\documentclass[compsoc,conference,a4paper,10pt,times]{IEEEtran}
\IEEEoverridecommandlockouts
\usepackage{cite}
\usepackage{amsmath,amssymb,amsfonts}
\usepackage{algorithmic}
\usepackage{graphicx}
\usepackage{textcomp}
\usepackage{bmpsize}
\usepackage{xcolor}
\usepackage{lipsum}
\usepackage{blindtext}
\usepackage{array}
\usepackage{multicol}
\usepackage{mwe}
\usepackage{comment}
\usepackage{array}
\newcolumntype{P}[1]{>{\raggedright\arraybackslash}p{#1}}
\usepackage[hyphens]{url}
\usepackage[colorlinks=true,urlcolor=black,hidelinks]{hyperref}
\def\BibTeX{{\rm B\kern-.05em{\sc i\kern-.025em b}\kern-.08em
    T\kern-.1667em\lower.7ex\hbox{E}\kern-.125emX}}
\hypersetup{breaklinks=true}
\begin{document}

\title{\#ISIS vs \#ActionCountersTerrorism: A Computational Analysis of Extremist and Counter-extremist Twitter Narratives}

\author{\IEEEauthorblockN{Fatima Zahrah}
\IEEEauthorblockA{\textit{Department of Computer Science} \\
\textit{University of Oxford}\\
Oxford, UK \\
fatima.zahrah@cs.ox.ac.uk}
\and
\IEEEauthorblockN{Jason R. C. Nurse}
\IEEEauthorblockA{\textit{School of Computing} \\
\textit{University of Kent}\\
Canterbury, UK \\
j.r.c.nurse@kent.ac.uk}
\and
\IEEEauthorblockN{Michael Goldsmith}
\IEEEauthorblockA{\textit{Department of Computer Science} \\
\textit{University of Oxford}\\
Oxford, UK \\
michael.goldsmith@cs.ox.ac.uk}
}

\maketitle

\begin{abstract}
The rapid expansion of cyberspace has greatly facilitated the strategic shift of traditional crimes to online platforms. This has included malicious actors, such as extremist organisations, making use of online networks to disseminate propaganda and incite violence through radicalising individuals. In this article, we seek to advance current research by exploring how supporters of extremist organisations craft and disseminate their content, and how posts from counter-extremism agencies compare to them. In particular, this study will apply computational techniques to analyse the narratives of various pro-extremist and counter-extremist Twitter accounts, and investigate how the psychological motivation behind the messages compares between pro-ISIS and counter-extremism narratives. Our findings show that pro-extremist accounts often use different strategies to disseminate content (such as the types of hashtags used) when compared to counter-extremist accounts across different types of organisations, including accounts of governments and NGOs. Through this study, we provide unique insights into both extremist and counter-extremist narratives on social media platforms. Furthermore, we define several avenues for discussion regarding the extent to which counter-messaging may be effective at diminishing the online influence of extremist and other criminal organisations.
\end{abstract}


\section{Introduction}
\label{introduction}
The past few decades have demonstrated how the Internet is playing an ever-increasing role in daily life, and has become an integral asset in society. In particular, the use of various digital technologies and online platforms for communication has been rapidly adopted into the home and work place alike. However, this has also introduced several implications as various malicious actors, or cyber-criminals, are quickly exploiting both the benefits afforded by such technologies as well as the vulnerabilities presented by them for their own criminal gains. Digital communities not only bring people closer together but also, inadvertently, provide criminals with new ways to access potential victims online. This has included extremist organisations strategically shifting their radicalisation and recruitment processes to online platforms for the purpose of indoctrinating individuals~\cite{Sabouni2017}. The same technologies that allow for a globalised world to interact seamlessly are also being utilised, adapted and abused by extremist organisations to target individuals and ensure organisational longevity \cite{Bertram2016TerrorismTI}. 

Shifting to social media and online means of communication has also provided the additional benefits of granting extremists with a perceived sense of anonymity, allowing access to an increased audience size, and utilising interactive features provided by online platforms to facilitate the acts of like-minded individuals exchanging radical thoughts \cite{Neo2016}. One of the leading examples of an extremist organisation making use of social media platforms for the purpose of radicalisation is ISIS. ISIS' early social media strategies on Twitter emphasised several of the unique characteristics of social media listed above. In many ways ISIS' highly energised recruitment efforts online and its reliance on the Internet have been central to its identity \cite{Greenberg2016}, in addition to introducing many initiatives from law enforcement agencies to monitor and remove offensive content online. The UK government specifically outlined online hate crime, with particular emphasis on online extremism, as one of the principal threats to cyber security in their National Cyber Security Strategy \cite{HMGovernmentUK2016}. Terrorist use of the Internet has also been highlighted as one of the primary forms of harmful and illegal content online in their Online Harms Paper \cite{HMGovernmentUK2019}.

Although the development of legislative and policing capabilities to prevent acts of extremism is clearly required, constructing approaches to reduce the radicalisation effects and impacts of extremist propaganda is also crucial to counter them. Such approaches are referred to as countering violent extremism (CVE), and have been perceived by both researchers and policy makers alike to being central to the process of addressing the pressing need to combat radicalisation to violence and extremism \cite{Davis2016}. CVE programs have generally included carrying focus groups and community engagement programs in particular demographics to encourage discussion around identity and social integration, and, specifically within the UK, the teaching of fundamental British values to deter extremism \cite{Webber2019}. More recently, the use of the Internet as an aid in CVE strategies has become increasingly apparent; for instance, some of the work carried out by Moonshot CVE, a social enterprise working to ``disrupt and eventually end violent extremism'' \cite{Moonshot}, makes use of Internet capabilities to develop counter-messaging campaigns and provide online interventions to vulnerable individuals. These are intended to carry out counter-extremism interventions through the same digital channels utilised by extremist groups, so as to reach the same vulnerable audiences. As social media becomes more present in daily life, CVE strategies must also embrace the same technologies to effectively discredit and nullify extremist groups \cite{Bertram2016TerrorismTI}.

Within the research landscape, several questions have been raised regarding the evaluation of such online CVE strategies, though very little has been carried out to explore how such initiatives compare against the content and strategies used by extremist organisations. Some studies have emphasised the influence that online messages can have on human behaviours and opinions. For instance, a study conducted by Frischlich et al. \cite{Frischlich2015} shows how people can be manipulated into agreeing with extremist viewpoints when under conditions of threat propagated by various media. Thus it can be assumed that online CVE could influence opinions in a similar way, however further research is required to strengthen this hypothesis.

In this article, we seek to advance current research by computationally exploring both extremist content found online, as well as the CVE strategy of counter-narratives \cite{Greenberg2016} designed to diminish the influence of extremist organisations on social media platforms. Specifically, we engage in a two part study that considers firstly, how extremists  and counter-extremist organisations craft content and secondly, how the psychological motivation behind the messages compares between them; thus, our contributions provide novel insight into both sides of online extremism. In particular, this study will apply computational techniques to analyse and compare the behaviour of various pro-extremist and counter-extremist Twitter accounts, and the effects this could have on influencing human behaviour. Through this research, we hope to understand the extent to which current CVE counter-extremist strategies on Twitter relate to extremist content, and identify potential avenues for future research on whether such CVE approaches can be made more effective.

The remainder of the report will be structured as follows. Section~\ref{relatedwork} will review the current literature on extremist use of the Internet. Section~\ref{methodology} will provide a detailed account of our approach and methodology, including the datasets and data analysis tools that were used. The results and observations from the analysis of both the pro-ISIS tweets and the counter-extremism tweets will be discussed in Section~\ref{results}. We then conclude and outline avenues for future work in Section~\ref{conclusion}.

\section{Related Work}
\label{relatedwork}
The phenomenon of radicalisation through online platforms has been researched extensively over the past decade by counter-terrorism and cyber security researchers. Many previous studies have examined key narratives incorporated by ISIS, as well as some of the major themes and components used within their online materials for the purpose of recruiting and disseminating propaganda. One such study is carried out by El-Badawy et al. \cite{EmmanEl-BadawyMiloComerford2015}, which closely examines violent Jihadi propaganda in order to understand their extremist ideology. The findings from this study showed that common beliefs shared by a majority of the global Muslim community, which may not necessarily be extreme, are frequently used to form solidarity with a wider target audience. Moreover, justifications from the Quran, Hadith or from scholarship are also often used to resonate with their Muslim audience \cite{EmmanEl-BadawyMiloComerford2015}. Similarly Torok \cite{Torok2013} provides a qualitative analysis of the social media accounts of a number of extremist groups; the results from this study identify a number of key discursive schemas, and highlight common themes used by various extremist Islamist groups, including `blaming of the West, unity of Islam, restoring the glory of Islam, and the embracing of death'. These findings also reinforce the observation that the unity of the wider Muslim community is a key radicalisation mechanism used to normalise extremist content and actions.

More similar to the research that will be covered in this paper, numerous studies have made use of computational approaches to analyse online extremist content and detect radicalisation. One such study is detailed by Vergani and Bliuc in \cite{Vergani2015}, where computational text-analysis tools were used to analyse the first 11 issues \textit{Dabiq} to investigate the evolution of ISIS' language. The results from this provided four key findings: affiliation or achievement plays a major role in motivating collective action of the group; ISIS is increasingly adopting emotional tones to increase influence, including anger and anxiety; ISIS texts exhibit more concern for women; and finally, they are making more use of Internet jargon to adapt itself to online environments and appeal to younger audiences. Another such study was conducted by Fernandez et al. in \cite{MiriamFernandezMoizzahAsif2018}, where they explored how online radical content could be detected, not just by searching for key terms and expressions associated with extremist discourse, but by further analysing the contextual semantics of such terms \cite{Fernandez2018}. This provided a more realistic and reliable radicalisation detection model by helping to discriminate radical content from content that only uses radical terminology, i.e. content simply reporting on events or sharing harmless religious rhetoric.

Despite the extensive research analysing online extremist content, to our knowledge, there have been few carried out to systematically or computationally analyse counter-extremism content currently existing online in a similar way. This could largely be due to the fact that, at present, there are few existing counter-extremism initiatives online, or at least few that exist on mainstream social media platforms such as Twitter. That being said, some of the work within this line of research includes a report by Ashour, which outlined a broad framework consisting of three major ``pillars'' that could be used to counter extremists narratives \cite{Ashour2010}. The first pillar was formed from a comprehensive message that dismantles and counter-argues against every dimension of the extremist narrative, such as the theological and political aspects. Secondly, choosing effective `messengers', who could be credible sources of information, namely former extremists who have been successfully de-radicalised, would also be imperative. Finally, the role of the media is essential to effectively disseminate counter-narrative content and attract a wider audience is also imperative. More recently, Wakeford and Smith~\cite{Wakeford2020} reinforce this point by arguing that it is not enough to simply delegitimise extremist posts; law enforcement agencies need to learn from extremist organisations, investigate and understand what makes them so influential, and harness this in their own counter-extremism efforts

The research detailed in this paper will therefore aim to fill the gap currently in this research landscape by providing more extensive empirical and statistical insight into the strategies used by pro-extremist users and how extremist content is constructed online. In particular, our work focuses on the extent to which the first and third pillars described above are currently being used in counter-extremist posts. We additionally bring some understanding into the psychological motivation behind their posts. By this, we specifically refer to how language can be manipulated to influence human behaviour. By comparing these findings to the content shared by counter-extremism agencies, we provide unique insight into both sides of the problem, and also provide avenues for discussion regarding the extent to which counter-messages may be effective at diminishing the online influence of extremist organisations.

\section{Methodology}
\label{methodology}
Our approach consists of analysing two datasets of tweets---one consisting of tweets from pro-ISIS accounts and the other consisting of tweets from counter-extremism agencies---to gain insight into the linguistic components used within them. We first use computational methods to carry out an empirical analysis to better understand the techniques used by pro-ISIS supporters and various counter-extremism organisations to promote their content. This will include comparing the usage of hashtags, links to external websites, and the most commonly used terms. Following this, we implement a more comprehensive linguistic analysis of the different sets of tweets to gain an understanding of how online narratives are framed. Through applying insights from previous research regarding the use of language and motivational theory (such as Regulatory Focus theory introduced by Higgins in \cite{Higgins1997}) in certain texts, this analysis will allow us to further explore how certain linguistic components can be used to influence behavioural change. This will also provide insight on whether online counter-narrative content can be crafted more effectively, for instance, by utilising more appropriate linguistic terms. Below, we describe the Twitter datasets that are used in the study, and the methods and tools used to analyse them.

\subsection{Datasets}
\label{datasets}
In order to analyse extremist content on social media, we acquired a publicly available dataset of noticeably pro-ISIS tweets posted by key ISIS-supporting Twitter accounts\footnote{https://www.kaggle.com/fifthtribe/how-isis-uses-twitter/data}. The dataset was published by the Kaggle data science community, and consists of over 17,000 English-only tweets retrieved from 112 distinct pro-ISIS supporter accounts over a period of three months during the aftermath of the November 2015 Paris terror attacks. These tweets were identified as being pro-ISIS after analysing specific indicators. This includes using certain key terms within their username, Twitter bio, or the actual tweet itself; following or being followed by other known radical accounts; or utilising images of ISIS logos or well-known radical leaders. This particular dataset has been used in several previous studies, including \cite{Fernandez2018} and \cite{Nouh2019}. Our study focussed on English-only tweets as the counter-extremist tweets used in this study were retrieved from English-speaking organisations only---further details of this are given below---making the analysis of the extremist and counter-extremist tweets more comparable.

Before using this dataset in our analysis, we first validated that these tweets were in fact posted by pro-ISIS Twitter accounts by manually checking the profiles of the 112 accounts using the Twitter API. Our assumption here is that, if the account no longer exists on Twitter or, in other words, has been blocked from the social media platform, then the account most likely did belong to an ISIS-supporting individual or group. This is due to the fact that the suspension or blocking of an account suggests that it had displayed malicious behaviour that did not comply with the Twitter terms of service. From this, we identified that only two of the Twitter accounts were not blocked and still existed on the platform at the time this research was conducted, where one of the accounts belonged to a journalist, and the other belonged to a researcher focusing on Jihadi groups. These two accounts and any tweets posted by these accounts were thus deleted from the pro-ISIS dataset, leaving a final total of 16,949 tweets from 110 pro-ISIS Twitter accounts.

In addition to the  pro-ISIS dataset, we used the Twitter API to retrieve a number of tweets from the Twitter accounts of major, English-speaking organisations specifically dealing with counter-extremism, including Governments and Law Enforcement Agencies (GLEAs), and NGOs. The accounts that tweets were retrieved from belong to the following agencies: The Commission for Countering Extremism in the UK (@CommissionCE); Counter Terrorism Policing UK (@TerrorismPolice); the UK Home Office (@ukhomeoffice); the US Department of State Bureau of Counterterrorism (@StateDeptCT); the Counter Extremism Project (@FightExtremism), an international policy organisation; the Global Center on Cooperative Security (@GlobalCtr), an international policy organisation; and Tech Against Terrorism (@techvsterrorism), an NGO supporting tech industries. These accounts were selected on the basis of the volume of tweets relevant to counter-extremism that were available to retrieve. 

Additionally, since the Twitter account of the UK Home Office does not solely post counter-extremism content, the tweets retrieved from it were filtered with the criteria that they include an extremism-related term (e.g., \textit{CVE}, \textit{terrorism}, \textit{extremist}). To gain deeper insight into how current counter narratives are constructed, we separated this collection of counter-extremism tweets into three further datasets: counter-extremism tweets from GLEAs in the UK, counter-extremism tweets from GLEAs in the US, and counter-extremism tweets from NGOs. Each of these datasets held between 2,000 and 3,000 tweets (with 2481 tweets from UK GLEAs, 2703 tweets from US GLEAs, and 2649 tweets from NGOs) and were analysed separately to investigate whether counter-narratives are crafted differently in each of the three organisational bodies. We then created a further dataset with the tweets from all the above mentioned counter-extremism datasets (with 7833 tweets in total) to easier compare the results from extremist and counter-extremist posts.

Before analysing the datasets, a series of pre-processing steps were carried out to clean the tweets and prepare them for further linguistic analysis. These steps included: (1) Removing any duplicate tweets or retweets from the datasets to reduce the levels of noise. (2) Removing all punctuation marks. (3) Removing any URLs from tweets. Similar data cleaning methods were used by \cite{Fernandez2018} and \cite{Nouh2019} prior to working with the dataset. It should also be noted here that account names and usernames were not used throughout the duration of this analysis---only the text used within the actual tweets were linguistically analysed after the specified Twitter accounts were chosen and organised into their appropriate sub-datasets.

\subsection{Analysis Framework and Methodology}
\label{framework}
Our analysis of the extremist and counter-extremist tweets is conducted with particular regards to two research questions:
\begin{itemize}
  \item RQ1: How are pro-extremist and counter-extremist messages constituted and what do they focus on promoting?
  \item RQ2: How do pro-extremist and counter-extremism Twitter accounts compare in terms of the methods used to disseminate content and the psychological motivation used within their tweets?
\end{itemize}
The initial empirical analysis will be used to examine data associated with each dataset of tweets, including the most commonly used hashtags and terms, and some of the `topics' that are associated with them. The results from this will then be compared across all five datasets of tweets to identify any similarities or differences between their respective approaches to promoting their content. This analysis will be carried out using the Pandas\footnote{https://pandas.pydata.org/} data analysis library and the `Natural Language Toolkit' (NLTK)\footnote{https://www.nltk.org/} provided by the Python programming language. 

To complement the aforementioned analysis, we use a framework majorly based on the theory of utilising regulatory focus to influence the thoughts and actions of a target audience through motivational regulation. In particular, this analysis is underpinned by the idea that regulatory focus distinguishes between two types of motivational regulation, promotion and prevention, as detailed by Higgins~\cite{Higgins1997}. Here, a promotion focus places emphasis on desires and potential goals, and often views these goals as hopes and aspirations \cite{Johnsen2014}. In contrast, a prevention focus places emphasis on potential losses, and tends to view goals as duties and obligations \cite{Higgins1997}. 

Moreover, the findings of Fuglestad et al. \cite{Fuglestad2008} also supported the notion that the regulatory focus and the frame of a message could be highly relevant to behavioural change, which they exemplified with smoking cessation and weight-loss interventions. Their study showed that a promotion focus could be related to the initiation of behavioural change (such as quitting smoking and dieting), while a prevention focus predicted the long-term maintenance of new, healthy behaviours. Some of these findings were applied to our study to observe if either prevention or promotion regulatory focus were used in the extremist or counter-extremist message frames to radicalise or de-radicalise target audiences respectively. This was assessed using an analysis framework based on the findings from Vaughn \cite{Vaughn2018}, which provided some statistical insight into the linguistic components used in both prevention and promotion regulatory focus.

To analyse the datasets of tweets, we used the programmatically coded dictionary from the Linguistic Inquiry and Word Count (LIWC) linguistic analysis tool to automate the process of extracting psychological meaning from textual content. A similar approach has been used in \cite{Vaughn2018} and other studies to examine and predict the psychological frames and behaviours of various groups as well as textual content, for instance, to predict depression \cite{Pennebaker2015}. LIWC is a widely used tool utilised in lexical approaches for personality measurement, and statistically analyses textual content based on 81 different categories by calculating the percentage of words in the input text that match predefined words in a given category. LIWC is used in our approach to assess the extent to which each of the datasets of tweets make use of promotion and prevention regulatory focus, in accordance with an analysis framework detailed by Vaughn in \cite{Vaughn2018}. Here, Vaughn specifies the LIWC categories that share significant differences in promotion-focussed text and prevention-focussed text. Further details of this have been provided in Section~\ref{liwc}. below. The next section will detail  the results from our analysis, and discuss the insights gained from this.

\section{Results and Discussion}
\label{results}
\subsection{Empirical Analysis}
\label{empircal analysis}

\begin{table*}[ht]
\caption{The 15 most used hashtags found in the pro-ISIS and counter-extremism tweets are summarised below.}
\centering
\begin{tabular}{| l | l | l | l | l |}
\hline
\textbf{Pro-ISIS Supp.} & \textbf{Counter-extremists} & \textbf{NGOs} & \textbf{US GLEAs} & \textbf{UK GLEAs}\\
\hline
\#isis--1577 & \#actioncountersterrorism--826 & \#cve--160 & \#counterterrorism--163 & \#actioncountersterrorism--826\\
\hline
\#syria--1373 & \#extremism--348 & \#isis--116 & \#isil--159 & \#extremism--304\\
\hline
\#is--677 & \#cve--255 & \#pve--96 & \#terrorist--102 & \#stop--97\\
\hline
\#iraq--634 & \#counterterrorism--238 & \#terrorism--84 & \#cve--95 & \#runhidetell--87\\
\hline
\#islamicstate--443 & \#isis--181 & \#counterterrosim--74 & \#terrorism--94 &  \#gunsoffourstreets--74\\
\hline
\#aleppo--406 & \#terrorism--179 & \#gifct--58 & \#hizballah--78 &  \#ctiru--67\\
\hline
\#amaqagency--332 & \#isil--163 & \#techvsterrorism--58 & \#isis--65 & \#extremists--54\\
\hline
\#breaking--324 & \#terrorist--135 & \#cft--54 & \#ct--65 & \#ath--51\\
\hline
\#russia--271 & \#pve--97 & \#aml--53 & \#syria--57 &  \#terrorists--47\\
\hline
\#breakingnews--252 & \#stop--97 & \#radicalization--50 & \#gctf--51 & \#ctaw2016pic--47\\
\hline
\#turkey--229 & \#runhidetell--87 & \#violentextremism--44 & \#bokoharam--50 & \#knowthegameplan--40\\
\hline
\#usa--216 & \#ct--85 & \#extremism--39 & \#iraq--39 & \#illegalguns--35\\
\hline
\#palmyra--215 & \#hizballah--78 & \#humanrights--37 & \#nigeria--37 & \#ctpolicingcareers--34\\
\hline
\#ypg--199 & \#syria--77 & \#technology--37 & \#iran--22 & \#worldcup--23\\
\hline
\#assad--159 & \#bokoharam--76 & \#un--31 & \#turkey--20 & \#besafebesound--19\\
\hline
\end{tabular}
\label{table:1}
\end{table*}

\subsubsection{Most Commonly Used Hashtags}
\label{hashtags}
Hashtags used by the pro-ISIS Twitter accounts followed most of the obvious political and extremist interests of radical Islamist as well as ISIS-specific adherents. In all the 16,494 pro-ISIS tweets, a total of 2418 distinct hashtags were detected, where 41\% of the tweets contained at least one hashtag. The 15 most used hashtags found in the pro-ISIS tweets---as well as those found in the other datasets of tweets---are summarised in Table~\ref{table:1}. The most popular hashtags by a wide margin were \textit{\#isis} and \textit{\#syria}, which were used 1577 and 1373 times respectively. Considering the fact that most ISIS-related activity was based in Syria and its surrounding areas, it is no surprise that a majority of the most common hashtags were related to locations where ISIS activity was most prevalent, including Iraq and Aleppo, as well as the states which had the most impact on ISIS activity---at the time of data collection---including Russia and the USA. The consistent usage of such hashtags helped amplify the coverage of ISIS-related news amongst supporter networks.

The tweets from counter-extremism NGOs used a total of 647 distinct hashtags (where 44\% of the tweets contained at least one hashtag), with the tweets from US GLEAs using a similar number of 605 distinct hashtags (where 57\% of the tweets contained at least one hashtag). However, the counter-extremism tweets from UK GLEAs used considerably fewer distinct hashtags, as only 375 unique hashtags were detected; however, we also found that hashtags were use significantly more often, where 76\% of the tweets contained at least one hashtag. This suggests these tweets were more consistent with their usage of hashtags to promote their content. In terms of those which were most used, similar strategies were used across all three counter-extremism datasets. As shown in Table~\ref{table:1}, the majority of the top hashtags used in the tweets were based around counter-terrorism and extremism (e.g., \textit{\#extremism}, \textit{\#cve}, \textit{\#terrorist}, \textit{\#counterterrorism}). 

A notable observation here is that the counter-extremism tweets from both the NGOs dataset and the US GLEAs dataset use the most similar hashtags, for instance, both datasets frequently use hashtags related to ISIS, including \textit{\#isis}, \textit{\#isil} and \textit{\#syria}. It should be noted here that the difference between the use of such hashtags by the pro-ISIS accounts and these counter-extremism accounts are that pro-ISIS tweets would use these hashtags to inform their audience of attacks and made by ISIS and to promote their cause, as quoted in the following tweet: \textit{``\#ISIS claims control in outskirts of south-\#Ramadi - 25 and Commander of 6th Regiment killed''}. US GLEAs would use such hashtags to inform on the US government's strategies on dealing with ISIS, as shown in the following tweet: \textit{``The US is dedicated to cutting off \#ISIL's financing, disrupting its plots \& stopping the flow of \#FTF's. \#CSISLive''}. This suggests that a majority of their counter-extremist policies were tailored to dealing with ISIS, since this is the only extremist organisation mentioned. NGOs would use hashtags relating to ISIS largely to promote research analysing ISIS activities and behaviours, for instance: \textit{``Successful terrorist operations have shifted from a 'tactical bonus' to 'strategic necessity' for \#ISIS, as 'online sphere has been tailored to facilitate these attacks more''.}

Contrastingly, the tweets from UK GLEAs did not make frequent use of ISIS-related hashtags. Instead, the most commonly used hashtags were concentrated around informing audiences on how to report acts of terrorism, including \textit{\#actioncountersterrorism}, \textit{\#stop}, \textit{\#runhidetell}, and \textit{\#knowthegameplan}. Another noteworthy point is that the top hashtags used by UK GLEAs were used more consistently, compared to the tweets from the other two datasets; the top hashtag in the UK dataset was used 826 times, which is considerably more that those used in the NGOs and US datasets, where the top hashtags were used 160 and 163 times respectively.

\subsubsection{Key Words and Topic Modelling}
\label{topics}
\begin{figure*}[ht]
\centering

    \includegraphics[width=0.35\linewidth]{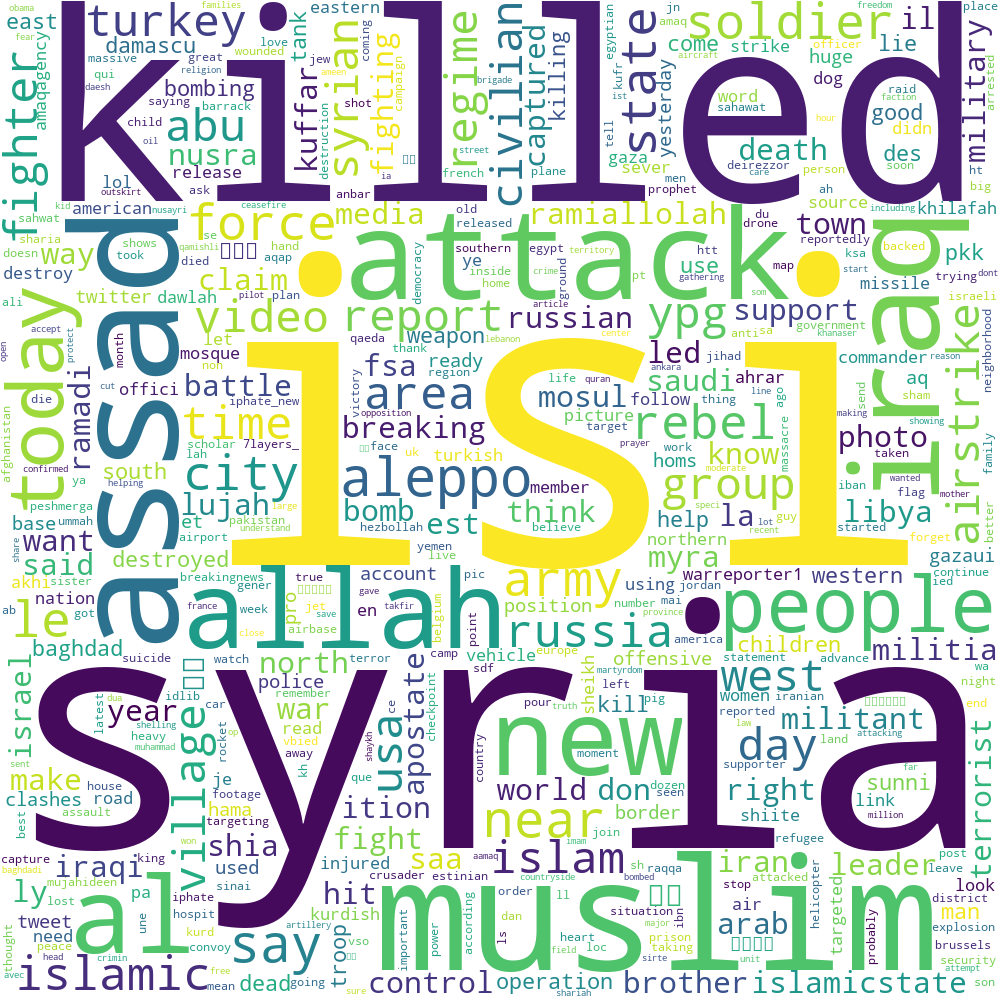}\hspace{1em}
    \includegraphics[width=0.35\linewidth]{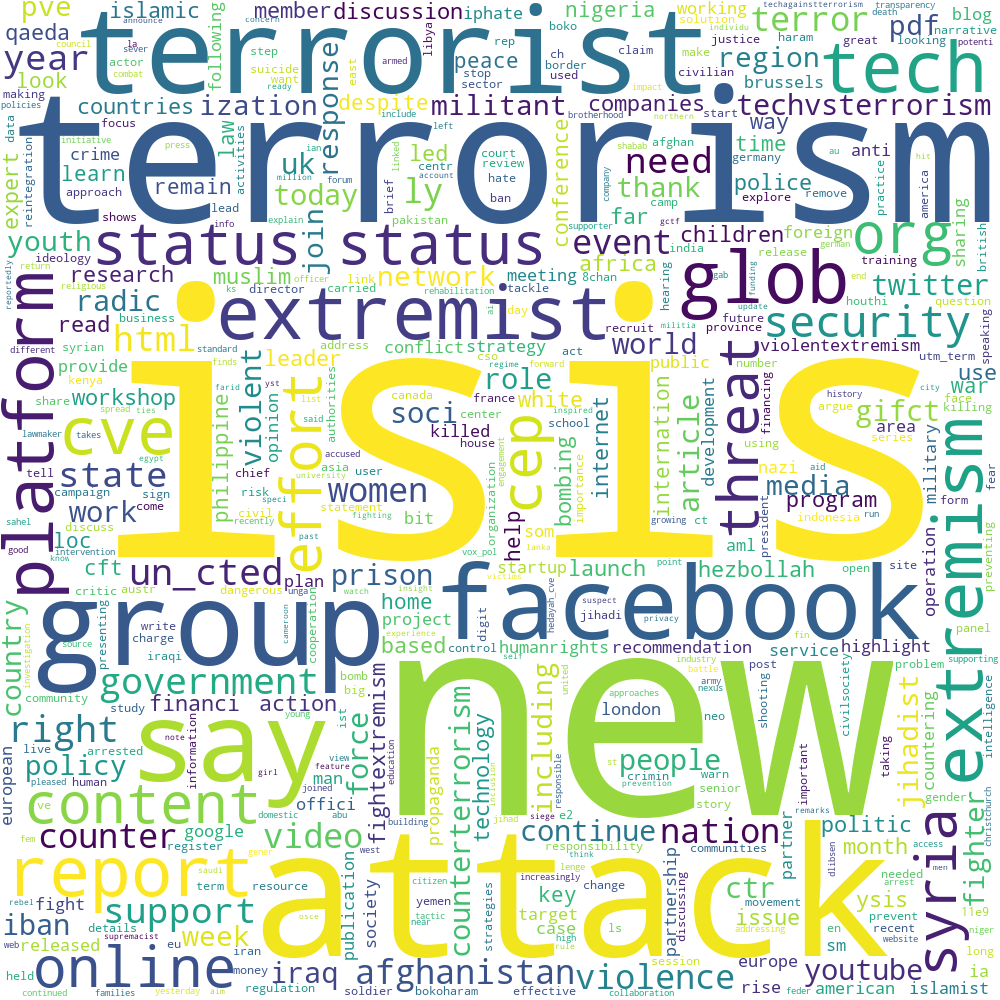}\par\medskip \medskip
    \includegraphics[width=0.35\linewidth]{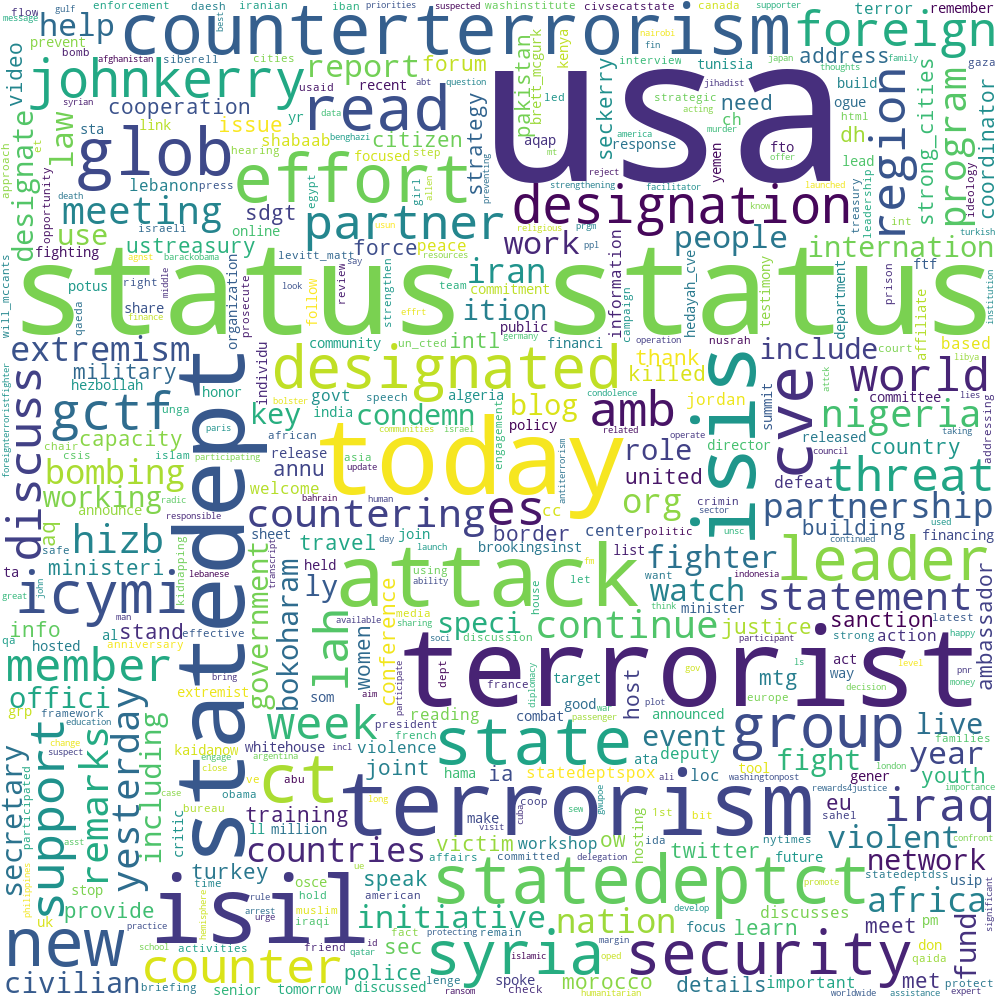}\hspace{1em}
    \includegraphics[width=0.35\linewidth]{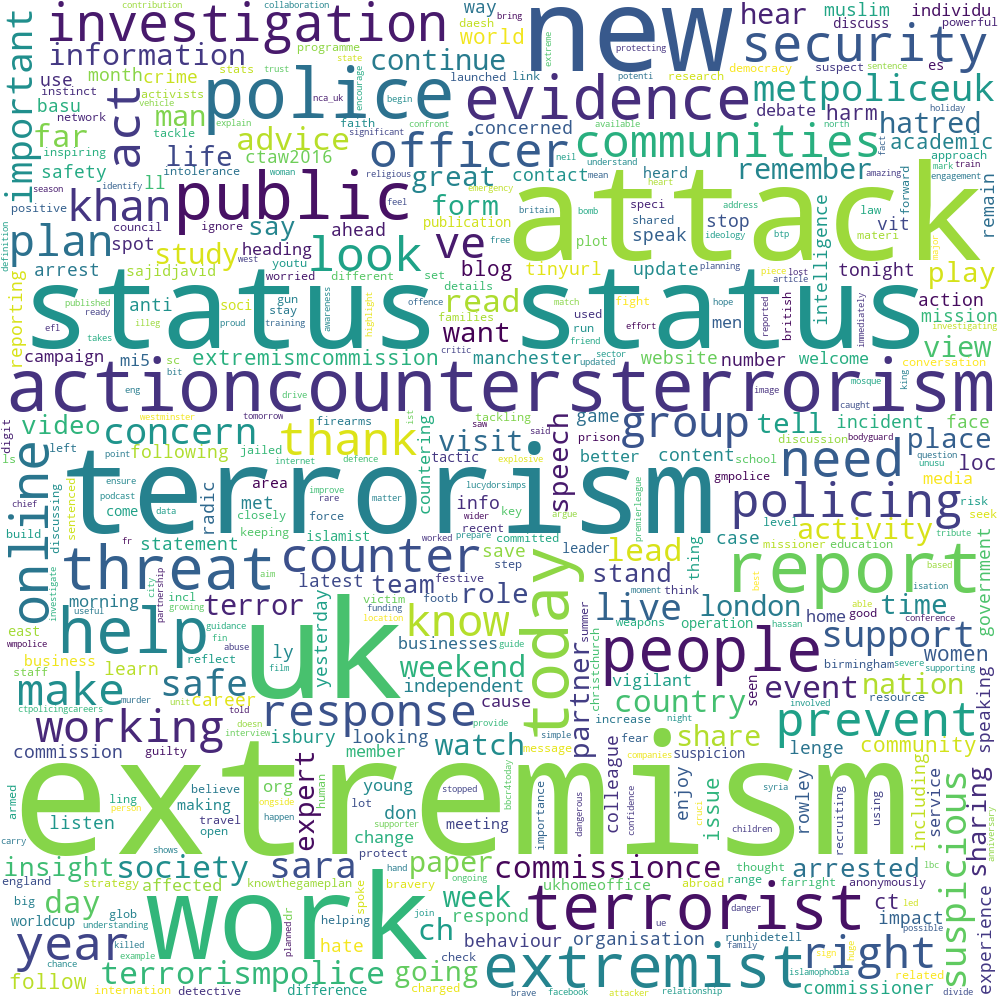}
\caption{Word clouds of most commonly used words in each dataset: pro-ISIS tweets (top-left), NGOs tweets (top-right), US GLEAs tweets (bottom-left), and UK GLEAs tweets (bottom-right).}
\label{fig:wordclouds}
\end{figure*}

The next part of the empirical analysis included determining which words and topics were mentioned the most in each dataset of tweets. The most used word in tweets from the pro-ISIS accounts was \textit{ISIS}, with \textit{Syria} being the second most common word. Along with the frequent mentioning of \textit{Aleppo}, \textit{Assad} and \textit{Iraq}, other common terms included \textit{killed}, \textit{army}, \textit{breaking}, \textit{soldiers} and \textit{attack}. Further details of the most common words in each dataset are provided in the word clouds in Figure~\ref{fig:wordclouds}. A topic model, using the Non-Negative Matrix Factorization (NMF) topic detection model, also provided useful insight into the most discussed subjects amongst pro-ISIS users as well as key emerging themes of ISIS ideologies, with the top 15 terms of each topic being detailed in Table~\ref{table:2}. We found that using the Non-Negative Matrix Factorization (NMF) topic detection model worked better with shorter texts, such as tweets, than other models, like Latent Dirichlet Allocation (LDA) \cite{godfrey2014}. A large proportion of the most discussed topics revolved around reporting the latest reports of attacks against ISIS as well as those instigated by ISIS. This includes Topic\#1, which seems to include discussions on Russia's and Turkey's involvement in ISIS territory in Syria (as reported in \cite{russia} and \cite{turkey}); as well as Topic\#5, seemingly discussing attacks from and on the US army during `The Battle of Mosul' (as reported in \cite{Ackerman2016} and \cite{Chulov2016}). Other topics heavily discussed the `fighters' of ISIS and their martyrdom, as well as prayers for `victory' and `reward' (Topic\#2 and Topic\#4), as shown in the following tweet: \textit{``Fighting Khawarij is greatest Jehad, whoever is killed by them receives the reward of a double martyr''}.

The most common words used in the counter-extremism tweets were similar across all three datasets, with the words \textit{terrorism}, \textit{extremism}, and \textit{counterextremism} being used most frequently in all sets of tweets. The counter-extremism tweets from both NGOs and US GLEAs also mentioned \textit{ISIS} or \textit{ISIL} on many occasions, whereas such terms were not amongst the most common words used by UK GLEAs. Tweets from UK GLEAs consistently made use of words relating to reporting extremist incidents including \textit{report}, \textit{police}, \textit{suspicious} and \textit{actioncountersterrorism}. In contrast, the tweets from NGOs and US GLEAs focussed more on informing about terrorist incidents and counter-extremism initiatives.
 
\begin{table*}[ht]

\centering
\caption{A topic model of the most discussed topics in each dataset.}
\begin{tabular}{| c | P{2.75cm} | P{2.75cm} | P{2.75cm} | P{2.75cm} | P{2.75cm} |}
\hline
 & \textbf{Pro-ISIS Supporters} & \textbf{Counter-extremists} & \textbf{NGOs} & \textbf{US GLEAs} & \textbf{UK GLEAs}\\
\hline
\textit{Topic\#1} & kill, soldier, today, airstrike, civilian, militant, wound, injure, bomb, Russian, children, yesterday, Turkish, dozen, Iraqi & Twitter, follow, status, find, visit, ISIS, use, social, media, discuss, internet, join, event, launch, watch & status, discuss, workshop, role, present, panel, participate, event, join, host, policies, brief, look, secure, first & icymi, yesterday, remark, ISIL, video, testimonies, destroy, strategies, coordinate, global, discuss, effort, envoy, countries, statement & extremist, content, terrorist, see, online, via, report, internet, social, media, act, remove, material, access, combat\\
\hline
\textit{Topic\#2} & islam, state, fighter, capture, force, via, unit, fight, group,  takfir, declare, call, war, muslim, martyredom & icymi, yesterday, remark, ISIL, video, testimonies, destroy, strategies, coordinate, global, discuss, effort, envoy, countries, statement & content, extremist, online, platform, facebook, media, social, white, youtube, group, nazi, video, remove,  supremacist, hate & design, terrorist, special, global, foreign, organise, individual, leader, yesterday, announce, member, entities, group, case, today & run, hide, tell, attack, safe, rare, advice, knife, gun, keep, simple, terror, prepare, weapon, firearm\\
\hline
\textit{Topic\#3} & Al Qaeda, sheikh, jabhat, leader, sham, jaish, release, ibn, jaysh, today, new, village, airstrike, Baghdadi, area & counter, terror, extreme, violent, police, UK, global, effort, right, discuss, prevent, support, coordinate, threat, ct & attack, kill, people, ISIS, bomb, Taliban, claim, Afghanistan, suicide, soldier, boko, haram, target, group, wound & read, latest, via, initial, program, article, policies, safe, remark, foreign, counterterror, recruit, prison, radical, challenge & report, suspicious, act, something, could, behaviour, anonymous, live, see, save, instinct, ignore, online, public, vigilant\\
\hline
\textit{Topic\#4} & Allah, may, accept, brother, protect, one, pleas, make, jazak, victorious, Muslim, sake, reward, love, bless & report, online, see, content, suspicious, extremist, help, act, via, terrorist, presence, active, visit, behaviour, find & report, new, recommend, juvenile, offend, policies, brief, violent, effort, rehabilitate, prevent, develop, need, societial, program & violent, counter, extreme, effort, fact, global, coalition, summit, prevent, build, support, extremist, local, terror, partner & presence, help, online, report, via, content, step, find, visit, us, button, speak, get, advice, website\\
\hline
\textit{Topic\#5} & ISI, US, Assad, fight, Muslim, support, rebel, Syrian, Mosul, help, want, group, back, Aleppo, YPG & run, hide, tell, safe, could, attack, rare, remember, simple, keep, knife, gun, terror, watch, weapon & counter, terror, global, extreme, forum, violent, internet, prevent, un, effort, strategies, threat, present, launch, nation &  attack, terrorist, condemn, statement, us, honor, victim, kill, remember, unit, bomb, year, die, mark, ago, families & game, secure, enjoy, plan, weekend, great, safe, time, address, stay, go, stadium, look, listen, check\\
\hline
\end{tabular}
\label{table:2}
\end{table*}

The topic model provided further insight into what content the counter-extremism tweets promoted in each of the three datasets. Again, tweets from both the NGOs and US GLEAs discussed similar topics. The majority of these topics were focussed on new \textit{counter extremism} efforts and policies, and reports of threats in various countries. Additionally, both sets of tweets discuss the activities of specific terrorist organisations, namely ISIS or ISIL, and strategies to counter them. The tweets from NGOs also  mentioned \textit{white supremacists} and their online presence, referring to other groups of extremists aside from radical Islamist organisations. Aside from this, NGOs would often promote workshops or events organised to discuss and promote counter-terrorism strategies and policies, hence, one of their frequently discussed topics was to promote tickets for such events, specifically Topic\#1 and Topic\#5. The tweets from US GLEAs would also refer to the US government and their response to terrorist incidents, as seen in Topic\#4 and Topic\#5.

In contrast, a huge majority of the counter-extremism tweets from UK GLEAs were concentrated around providing advice on how to respond to terrorist incidents as well as promoting the appropriate channels for reporting such incidents. In this way, the tweets were addressed to an audience from a more specific demographic, i.e., UK residents who could report incidents to appropriate law enforcement departments within the UK. Moreover, tweets from this dataset also promoted campaigns against violent crimes in general, such as gun and knife crime in Topic\#2, and did not refer to any specific form of extremism or organisation, for instance: \textit{``our advice to the public on what do if caught in a gun or knife terror attack. It could keep you, your friends and family safe \#ActionCountersTerrorism''}. Such tweets were often posted multiple times, showing that UK GLEAs often repeated tweets to emphasise its importance to their target audience.

\subsection{LIWC Analysis}
\label{liwc}
\subsubsection{Exploring Motivational Theory}
Although there are 81 categories in the LIWC standard dictionary, only the categories that were proven in \cite{Vaughn2018} to indicate the use of a regulatory focus were used in our analysis framework, though some observations were also made for LIWC categories which showed notable differences between the datasets. Table~\ref{table:3} summarises the results from the LIWC analysis whilst assessing the extent to which a promotion or prevention focus were used. The table shows the mean percentage of all the words used within the tweets  that fall into a particular LIWC category. For instance, a mean percentage of 1.35 for \textit{positive emotion} words implies that 1.35\% of the words used within the respective dataset were associated with positive emotion. Example words of each LIWC category are also included in the table.

Overall, counter-extremism tweets make use of a promotion focus more than the pro-ISIS tweets. Counter-extremism content tended to use the most language associated with \textit{positive emotion}, which was specified as an indicator for descriptions of pursuing hopes, and therefore associated with a promotion focus. Similarly, counter-extremism tweets made use of words related to \textit{work}, \textit{achievement} and \textit{leisure} more than pro-ISIS tweets, which also indicated a stronger promotion focus. Further still, such tweets from UK GLEAs had a stronger promotion focus than any of the other counter-extremism organisations. The results from the LIWC analysis also clearly show that the pro-ISIS tweets had a very small mean percentage for the number of words used that were associated with any of the four defining conditions for content with a promotion focus. This suggests that extremists do not tend to frame their messages around promotion, or view their goals as hopes and aspirations.

When looking at the results for the prevention focus section of the LIWC analysis, we can observe, interestingly, that counter-extremism tweets posted by UK GLEAs also made use of language associated with a prevention focus more than any of the other sets of tweets. The results for the UK GLEAs showed a greater mean percentage for almost all of the distinctive LIWC categories indicating the strong use of prevention-focussed content. Pennebaker's findings in \cite{pennebakersecret} found that, when compared to descriptions of pursuing hopes, descriptions of pursuing duties were more likely to include stories about dynamic social interactions and processes. This also infers that a higher percentage of function words including \textit{pronouns}, \textit{prepositions}, \textit{auxiliary verbs}, \textit{negations}, and \textit{conjunctions} are used in the text, which can be observed from Table~\ref{table:3}.

\begingroup
\begin{table*}[ht]

\centering
\caption{Results from the LIWC analysis when observing regulatory focus.}
\begin{tabular}{|m{5em}|c | c | c | c | c | c |}
\hline
\textbf{LIWC Categories} & \textbf{Examples} & \textbf{Pro-ISIS Supporters} & \textbf{Counter-extremists} & \textbf{NGOs} & \textbf{US GLEAs} & \textbf{UK GLEAs}\\
\hline 
\multicolumn{7}{|l|}{\textbf{Promotion Focus}} \\
\hline 
\textit{Positive Emotion} & happy, pretty, good & 1.35 & 2.46 & 2.24 & 2.22 & 3.78\\
\hline
\textit{Work} & work, class, boss & 1.08 & 5.68 & 4.08 & 3.75 & 4.72\\
\hline
\textit{Achievement} & try, goal, win & 0.97 & 1.69 & 2.04 & 2.50 & 2.23\\
\hline
\textit{Leisure} & house, TV, music & 0.57 & 2.57 & 0.82 & 1.08 & 1.36\\
\hline
\multicolumn{7}{|l|}{\textbf{Prevention Focus}} \\
\hline
\textit{Function words} & it, to, no, very & 23.97 & 27.35 & 26.77 & 25.95 & 37.10\\
\hline
\textit{Pronouns} & I, them, itself & 4.29 & 5.04 & 2.98 & 3.65 & 8.79\\
\hline
\textit{Personal Pronouns} & I, them, her & 2.61 & 2.88 & 1.54 & 2.06 & 5.14\\
\hline
\textit{Conjunctions} & but, whereas & 2.06 & 3.11 & 2.67 & 2.30 & 4.51\\
\hline
\textit{Negations} & no, never, not & 0.64 & 0.32 & 0.20 & 0.26 & 0.53\\
\hline
\textit{Negative Emotion} & hate, worthless, enemy & 2.63 & 3.60 & 3.60 & 3.70 & 4.08\\
\hline
\textit{Social Processes} & talk, us, friend & 4.95 & 9.24 & 5.78 & 5.53 & 11.14\\
\hline
\textit{Family} & mom, brother, cousin & 0.24 & 0.13 & 0.16 & 0.12 & 0.11\\
\hline
\textit{Friends} & pal, buddy, coworker & 0.06 & 0.23 & 0.27 & 0.30 & 0.17\\
\hline
\end{tabular}
\label{table:3}
\end{table*}
\endgroup

In addition to this, messages from a prevention focus tend to focus on the avoidance of negative outcomes, and because of this, often use more language associated with \textit{negative emotions}. The results show that UK GLEA tweets had a higher percentage for this particular LIWC category as well, supporting the observation that it made use of a prevention focus the most compared to the other datasets. On the other hand, the pro-ISIS tweets generally seemed to use prevention-focussed narratives less than counter-extremism tweets, and so did not put as much emphasis on duties and obligations as the counter-extremism tweets did.

Overall, the results from this linguistic analysis show that the pro-ISIS tweets used much less regulatory focus, whether promotion or prevention, in their narratives than the counter-extremism tweets. Further analysis would therefore be required to assess their radicalisation techniques. However, an observation that is common across the results for all five datasets is that they all, generally, used a prevention focus in their messages than a promotion focus. This could suggest that both extremist and counter-extremist narratives view their goals more as duties and obligations than hopes and aspirations. Through knowledge gained from previous studies such as \cite{Higgins1997} and \cite{Johnsen2014}, we can infer that using such a focus could be useful to maintain behavioural change, though it does not necessarily inspire initial behavioural change as narratives from a promotion focus would do. Due to this, it could be beneficial for online counter-extremism narratives to make use of more promotion-focussed content and pursue positive end-states or goals in order to initiate de-radicalisation, although prevention-focussed narratives would still be necessary to maintain these efforts.

\subsubsection{Additional Observations}
\label{additional obvs}
In addition to exploring the usage of regulatory focus, LIWC was also used to analyse each of the datasets for any other notable distinctions in the linguistic composition of the tweets. Significant observations have been summarised in Table~\ref{table:4}. Whilst conducting this analysis, an immediate distinction that can be seen is the usage of pronouns in each dataset. Overall, all of the counter-extremism datasets generally used less singular first-person pronouns (such as \textit{I}, \textit{me}, \textit{my}), and more plural first-person pronouns (such as \textit{we}, \textit{our}, \textit{us}) than the pro-ISIS tweets. Second-person pronouns (such as \textit{you}, \textit{yours}, \textit{yourself}) were present in the tweets from UK GLEAs significantly more than any of the other datasets of tweets. This is in line with the previous observations made from the empirical analysis where the counter-extremism tweets from UK GLEAs mainly addressed their audience directly to inform them of how to properly report and protect against terrorist incidents. When looking at third-person pronouns (such as \textit{she}, \textit{he}, \textit{they}), the results from the LIWC analysis show that they were used more in the tweets from the pro-ISIS accounts than any of the counter-extremism tweets.

The use of pronouns in speech and text has been studied extensively in previous works, and has often been identified as a discursive tool used to persuade audiences. This effect of persuasion is partly due to the variability of the scope of reference of the pronouns used, which is determined by the audience, who can then interpret whether they are inclusive or exclusive of them \cite{Wilson1990, Zupnik1994}. In particular, the use of personal pronouns such as \textit{we}, \textit{you}, \textit{our} and \textit{us} is a common persuasive technique used in writing to make audiences feel more immediately involved. The LIWC analysis shows that this particular strategy is used more in the counter-extremism tweets than the pro-ISIS tweets; more specifically, the tweets from the UK GLEAs used such pronouns significantly more than the other sets of tweets. However, it should be noted here that the pro-ISIS tweets used such second-person pronouns (e.g. \textit{you}, \textit{yours}) more than the counter-extremism tweets from NGOs and US GLEAs. 

\begin{table*}[ht]

\centering
\caption{Additional observations made from the LIWC analysis.}
\begin{tabular}{| c | c | c | c | c | c |}
\hline
\textbf{LIWC Category} & \textbf{Pro-ISIS Supporters} & \textbf{Counter-extremists} & \textbf{NGOs} & \textbf{US GLEAs} & \textbf{UK GLEAs}\\
\hline
\textit{I} & 0.50 & 0.08 & 0.06 & 0.09 & 0.15\\
\hline
\textit{We} & 0.46 & 1.40 & 0.78 & 1.41 & 2.08\\
\hline
\textit{You} & 0.52 & 0.94 & 0.19 & 0.19 & 2.38\\
\hline
\textit{She/he} & 0.42 & 0.15 & 0.15 & 0.11 & 0.21\\
\hline
\textit{They} & 0.71 & 0.31 & 0.35 & 0.26 & 0.38\\
\hline
\textit{Anxiety} & 0.25 & 1.79 & 1.42 & 1.60 & 2.13\\
\hline
\textit{Religion} & 1.30 & 0.43 & 0.57 & 0.40 & 0.27\\
\hline
\textit{Death} & 0.85 & 0.25 & 0.40 & 0.27 & 0.07\\
\hline
\end{tabular}
\label{table:4}
\end{table*}

Another noteworthy point here is that making use of third-person pronouns in political discourse is a tactic that can be used to delineate the level of commitment and involvement of an organisation to the statement being made \cite{Ho2013}. This is used most in the pro-ISIS tweets, largely due to the fact that most of these tweets are from pro-ISIS supporters, and likely not ISIS themselves, though the counter-extremism tweets---especially those from GLEAs---were directly from official representatives of the organisations. This shows that, in general, the counter-extremism agencies were more involved or committed to any future responsibilities declared in the tweets.

The results from the LIWC analysis also showed significant differences in the use of language associated with anxiety (e.g. \textit{nervous}, \textit{afraid}, \textit{tense}). Generally, the counter-extremism tweets used more anxiety-related language than the pro-ISIS tweets from UK GLEAs using such language the most, which is supported by the findings from Vergani and Bliuc in \cite{Vergani2015}. The use of language related to death (e.g. \textit{kill}, \textit{bury}, \textit{grave}) was more common in the pro-ISIS tweets, though this is justifiable considering our analysis showed that themes of martyrdom and the attacks on ISIS, as well as the findings from Torok in \cite{Torok2013}, were frequently discussed. Another observation is that the pro-ISIS tweets used more religion-associated language than the counter-extremism tweets. Tweets from US GLEAs and NGOs referred to religion slightly more than those from UK law GLEAs, though this could largely be due to the fact that these datasets were shown to discuss ISIS frequently, as noted earlier in Section~\ref{empircal analysis}. Recent research, such as the study carried out by El-Said in \cite{El-Said2012}, has shown that a major de-radicalisation strategy, particularly when countering ISIS narratives, is to involve clerics and scholars to promote authentic religious teachings, and use them to refute misinformed religious teachings propagated by extremists. This provides a further area of development for counter-extremism campaigns on social media platforms, where making use of such religious teachings could help to directly counter a significant amount of extremist content online.

\subsection{Limitations}
\label{limitations}
Despite gaining some useful insights through this study, certain limitations of our approach could have impacted our observations. The first limitation that affects any research carried out in counter-extremism is that it is very difficult to ethically measure the effectiveness of counter-extremism initiatives. This makes it challenging to come to any concrete conclusions about how such counter-narratives can be improved since there is no efficient way to evaluate them (other than first hand experience with individuals impacted). Additionally, measuring the effect of online extremist or counter-extremist content on behavioural change is also hard to do with ethically-sound methodologies, and therefore can mainly be supported by the findings from previous studies and research, as done in this paper. However, since it is undeniable that online extremists played a major role in the radicalisation of their target audiences on mainstream social media, it should be possible for counter-extremism narratives to reach the same platforms as online extremists, and therefore distribute content that is accessible and influential among their target audiences \cite{Ashour2009}.

Another point to note about is that the counter-extremism tweets were gathered from different time frames than the pro-ISIS tweets. For instance, tweets from UK GLEAs were collected from October 2016 to September 2019, tweets from US GLEAs were collected from March 2013 to September 2019, and tweets from NGOs were collected from January 2015 to September 2019. This wide time span is largely due to the lack of counter-extremism content available on Twitter (an interesting observation in itself). In our study, we felt that it was more important to gain a dataset of tweets large enough to analyse and compare with the results of the dataset of the pro-ISIS tweets (which were all from 2015 and 2016). It should be noted here, however, that most of the counter-extremism accounts started posting more frequently at around the same time the pro-ISIS tweets were posted (in 2015 and 2016), which is when ISIS supporters were more prevalent on Twitter \cite{Ceron}.

\section{Conclusions and Future Work}
\label{conclusion}
Up until recently, regulation of the internet against organisational crime and extremism in online spaces has mainly concentrated on disruption efforts. Ultimately, our work suggests that perhaps other initiatives, such as CVE, could be used to combat online radicalisation. From this study, we sought to advance the current research in online online extremism and counter-extremism narratives by comparing the online behaviours of Twitter accounts from both extremist and counter-extremist organisations, and assessing how the two sets of messages compare with each other. To our knowledge, this is the first work to explicitly compare extremist and counter-extremist content in this way, whilst also applying psychological motivational theory to explore how such posts can influence behavioural change in online audiences. Although our study analyses data from one particular use-case of online radicalisation through pro-ISIS tweets, we believe that a similar methodology could be applied to other use cases of radicalisation using online platforms to gain further insights into the effectiveness of counter-extremism strategies.

Through performing linguistic analysis on datasets of tweets from pro-ISIS supporters and various counter-extremism organisations, we found that, oftentimes, counter-extremism tweets from certain agencies---namely US GLEAs and NGOs---would promote topics and use hashtags which were also used frequently by pro-ISIS supporters. This included frequent discussion around ISIS activity and use of the hashtag \textit{\#ISIS}. In contrast, counter-extremism tweets from UK GLEAs seemed to share completely different content when compared to each of the other datasets of tweets. In this case, the majority of their posts were crafted for the purpose of informing online audiences on how to report or protect themselves against possible extremist activity, where specific extremist groups were rarely referred to. Consequently, most of these posts were not constructed to directly counter extremist content being posted online.

In terms of the psychological motivation behind the tweets, with specific regards to Regulatory Focus Theory, we found that counter-extremism tweets  generally seemed to use regulatory focus more than the pro-ISIS tweets, with tweets from UK GLEAs using such motivational theory the most. An avenue for future work here would be to analyse the pro-ISIS tweets with more advanced methods, frameworks or tools to assess any other radicalisation techniques used by extremists to radicalise and manipulate their audience. Our findings also showed that, overall, both extremist and counter-extremist tweets used prevention-focussed narratives more than promotion-focussed narratives. An area for further study would therefore be to explore whether using more promotion-focussed narratives would be an effective counter-extremism strategy. 

Previous research conducted by Higgins in \cite{Higgins1997} suggested that promotion-focussed narratives inspired initial behavioural change, whereas using a prevention focus could facilitate the maintenance of this behavioural change. Thus, another hypothesis that may be worth investigating here would be whether the regulatory focus of such online extremism and counter-extremism content changed chronologically; did such narratives make use of a promotion focus initially, and then shift to using a prevention focus? Such theories could also be explored on other online platforms, not just Twitter.

\bibliographystyle{IEEEtran}
\bibliography{references.bib}

\end{document}